\documentclass[a4paper]{jpconf}
\usepackage{graphicx}
\usepackage[square,sort&compress,numbers]{natbib}
\usepackage[colorlinks=true,urlcolor=blue,anchorcolor=blue,citecolor=blue,filecolor=blue,linkcolor=blue,menucolor=blue,linktocpage=true,unicode=true,bookmarksopen=true,pdfa=true]{hyperref}
 
\usepackage{amsmath}
\usepackage{amssymb}
\usepackage{slashed}
\usepackage{mathtools}
\usepackage{lmodern}
\usepackage[T1]{fontenc}
\usepackage{bbm}
\DeclareMathAlphabet{\mathbfi}{OML}{cmm}{b}{it}
\usepackage{color}
\usepackage{caption}
\usepackage{subcaption}

\let\originalleft\left
\let\originalright\right
\renewcommand{\left}{\mathopen{}\mathclose\bgroup\originalleft}
\renewcommand{\right}{\aftergroup\egroup\originalright}

\makeatletter
\newenvironment{equations}[1][]{\subequations\ifx\relax#1\relax\else\label{#1}\fi\align\ignorespaces}{\endalign\ignorespacesafterend\endsubequations}
\def\@spliteq#1{\begin{equation}\begin{split}#1\end{split}\end{equation}}
\def\splitequation{\collect@body\@spliteq}

\makeatother

\renewcommand{\vec}[1]{{\ifnum9<1#1\mathbf{#1}\else\ifcat\noexpand#1\relax\boldsymbol{#1}\else\mathbfi{#1}\fi\fi}}
\newcommand{\mathe}{\mathrm{e}}
\newcommand{\mathi}{\mathrm{i}}
\let\oldre\Re
\let\oldim\Im
\renewcommand{\Re}{\oldre\mathfrak{e}\,}
\renewcommand{\Im}{\oldim\mathfrak{m}\,}
\newcommand{\total}{\mathop{}\!\mathrm{d}}

\newcommand{\1}{\mathbbm{1}}
\newcommand{\eqend}[1]{\,#1}
\newcommand{\bigo}[1]{\mathcal{O}\left({#1}\right)}

\makeatletter
\def\expect{\@ifnextchar[{\@expecttw@}{\@expect@ne}}
\def\@expecttw@[#1]#2{\left\langle{#2}\right\rangle^\text{#1}}
\def\@expect@ne#1{\left\langle{#1}\right\rangle}
\makeatother

\newcommand{\diver}{\mathcal{N}}

\def\mailto#1{\href{mailto:#1}{\texttt{#1}}}

\usepackage{tikz}
\usepackage{tikz-feynman}

\begin{document}
\title{Trace anomalies for Weyl fermions: too odd to be true?}

\author{S~Abdallah\ad{1}, S~A~Franchino-Vi{\~n}as\ad{2} and M~B~Fr{\"o}b\ad{3}}

\address{\ad{1} Fakult{\"a}t f{\"u}r Mathematik, Universit{\"a}t Regensburg, Universit{\"a}tsstra{\ss}e 31,\\\hspace{8pt}93053 Regensburg, Germany.}

\address{\ad{2} Helmholtz-Zentrum Dresden-Rossendorf, Bautzner Landstraße 400,\\\hspace{8pt}01328 Dresden, Germany.}

\address{\ad{3} Institut f{\"u}r Theoretische Physik, Universit{\"a}t Leipzig, Br{\"u}derstra{\ss}e 16,\\\hspace{8pt}04103 Leipzig, Germany.}

\vspace*{5pt}
\address{E-mail: \mailto{sami.abdallah@mathematik.uni-regensburg.de}, \mailto{s.franchino-vinas@hzdr.de}, \mailto{mfroeb@itp.uni-leipzig.de}}

\begin{abstract}
We review recent discussions regarding the parity-odd contribution to the trace anomaly of a chiral fermion. We pay special attention to the perturbative approach in terms of Feynman diagrams, comparing in detail the results obtained using dimensional regularization and the Breitenlohner--Maison prescription with other approaches.
\end{abstract}

\section{Introduction}
\label{sec:introduction}
Symmetries in physics are generally well-received, inasmuch as they usually lead to great simplifications. In classical theories, we are used to studying symmetries at the level of the action: a symmetry is present if the action remains invariant when the fields are transformed accordingly. Upon the quantization process, such symmetries may survive or be broken; in the latter case, one usually speaks of the appearance of anomalies. Notably, anomalies are as resistant and as weak as spiderwebs: the former is a consequence of their cohomological foundations; the latter is reflected in the fact that, since their computation involves several subtleties, one could easily make the would-be anomalies vanish into thin air with an inconsistent computation.

For general insight into this topic, we refer to the lectures~\cite{Bilal:2008qx,Harvey:2005it}, as well as the foundational articles~\cite{Adler:1969gk, Bell:1969ts, Alvarez-Gaume:1983ihn,AlvarezGaume:1983cs}. Since their very discovery, it has been patent that anomalies have profound physical implications~\cite{Bell:1969ts}. In recent years we have seen a renewed interest in them, based on diverse grounds. On the one hand, their consistency in the effective field theory framework of the standard model~\cite{Feruglio:2020kfq,Passarino:2021uxa} and beyond the standard model~\cite{Nakagawa:2021wqh,Nakayama:2018dig} has been studied. On the other hand, anomalies have several consequences for the behaviour of condensed matter systems~\cite{Chernodub:2021nff}; as an example, the gravitational contribution to the axial anomaly has a measurable effect on transport phenomena~\cite{Gooth:2017mbd}. A third motivation for their study is the close relation between trace (also called Weyl) anomalies and conformal field theories~\cite{Nakayama:2012gu,Coriano:2018bsy, Bzowski:2020kfw, Coriano:2021nvn, Chalabi:2021jud}, together with the preponderance acquired by the latter in the realm of the AdS/CFT correspondence.

The invariance of a field theory under local Weyl transformations classically implies the on-shell tracelessness of the energy-momentum (EM) tensor $T^{\mu\nu}$. At the quantum level, the EM tensor becomes an operator, whose renormalized expectation value may develop a nonvanishing value; in such a case, it is said that we are in the presence of a trace anomaly. First found by Capper and Duff~\cite{Capper:1974ic, Duff:2020dqb}, on dimensional grounds and requiring covariance, the structure of the trace anomaly in four dimensions is constrained to be of the form
\begin{equation}
\label{eq:anomaly_expression}
g_{\mu\nu} \expect{T^{\mu\nu}}_\text{ren} = w C^{\mu\nu\rho\sigma} C_{\mu\nu\rho\sigma} + b \mathcal{E}_4 + c \nabla^2 R + f \epsilon^{\mu\nu\rho\sigma} R_{\mu\nu\alpha\beta} R_{\rho\sigma}{}^{\alpha\beta} \eqend{,}
\end{equation}
where $w$, $b$, $c$ and $f$ are numerical coefficients, $R^{\mu\nu\rho\sigma}$ is the Riemann tensor, $R$ is the Ricci scalar and $\nabla^2$ is the d'Alembertian. 
In terms of the Riemann tensor, the Ricci tensor and the Ricci scalar, the Weyl tensor $C_{\mu\nu\rho\sigma}$ and the Euler density $\mathcal{E}_4$ can be written as
\begin{equations}
C^{\mu\nu\rho\sigma} C_{\mu\nu\rho\sigma} &= R^{\mu\nu\rho\sigma} R_{\mu\nu\rho\sigma} - 2 R^{\mu\nu} R_{\mu\nu} + \frac{1}{3} R^2 \eqend{,} \\
\mathcal{E}_{4} &= R^{\mu\nu\rho\sigma} R_{\mu\nu\rho\sigma} - 4 R^{\mu\nu} R_{\mu\nu} + R^2 \eqend{.}
\end{equations}
Under parity, all terms in the trace anomaly in Eq.~\eqref{eq:anomaly_expression} are invariant except for the last one, which changes sign. Therefore, assuming that the theory is CPT invariant as any relativistic quantum field theory must be, the coefficients $w$, $b$, and $c$ of the parity-even terms must be real, while the coefficient $f$ of the parity-odd term (the Pontryagin density) must be purely imaginary.

An intuitive way to see the appearance of the trace anomaly is as follows: Consider a four-dimensional quantum field theory on curved space whose classical action $S$ is Weyl-invariant. If divergences arise in the computation of the effective action $\Gamma$, they comprise local terms $\Delta \Gamma$, which are expected to be also Weyl-invariant. This property can be seen for example from cohomological arguments~\cite{Bonora:1983ff} or using spectral techniques~\cite{DeWitt:2003}. Therefore, in dimensional regularization the parity-even contribution $\Delta \Gamma$ is constrained to be of the form
\begin{equation}
\Delta \Gamma = \frac{\mu^{n-4}}{n-4} \int \left( a_1 C^{\mu\nu\rho\sigma} C_{\mu\nu\rho\sigma} + a_2 \mathcal{E}_4 \right) \sqrt{-g} \total^n x \eqend{,}
\end{equation}
where $a_1$ and $a_2$ are coefficients which are fixed by the requirement that the effective action be finite~\cite{Duff:1977ay}.
Now a straightforward computation shows that
\begin{equations}
\frac{2}{\sqrt{- g}} g_{\mu\nu} \frac{\delta}{\delta g_{\mu\nu}} \int C^{\mu\nu\rho\sigma} C_{\mu\nu\rho\sigma} \sqrt{-g} \total^n x &= (n-4) \Big( C^{\mu\nu\rho\sigma} C_{\mu\nu\rho\sigma} + \frac{2}{3} \nabla^2 R \Big) \eqend{,} \\
\frac{2}{\sqrt{-g}} g_{\mu\nu} \frac{\delta}{\delta g_{\mu\nu}} \int \mathcal{E}_4 \sqrt{-g} \total^n x &= (n-4) \mathcal{E}_4 \eqend{;}
\end{equations}
thus, due to the pole in $\Delta \Gamma$, the trace of the EM tensor computed from $\Gamma + \Delta \Gamma$ acquires the anomalous contribution
\begin{equation}
\frac{2}{\sqrt{-g}} g_{\mu\nu} \frac{\delta}{\delta g_{\mu\nu}} \Delta \Gamma = a_1 \Big( C^{\mu\nu\rho\sigma} C_{\mu\nu\rho\sigma} + \frac{2}{3} \nabla^2 R \Big) + a_2 \mathcal{E}_4 \eqend{.}
\end{equation}

The coefficients $w$, $b$, $c$ and $f$ have been computed in several cases. In a theory with $N_s$ free scalar fields, $N_f$ free Dirac fermions and $N_m$ free Maxwell fields, they are given by\footnote{Notice that in principle one can obtain a different value of $c$ by introducing a finite counterterm proportional to $R^2$; here we follow the convention in Ref.~\cite{Duff:1977ay}.}~\cite{Duff:1977ay}
\begin{equations}[eq:coefficients_anomaly]
w &= \frac{1}{120 (4\pi)^2} \left( N_s + 6 N_f + 12 N_m \right) \eqend{,} \\
b &= - \frac{1}{360 (4\pi)^2} \left( N_s + 11 N_f + 62 N_m \right) \eqend{,} \\
c &= \frac{2}{3} w \eqend{,} \\
f &= 0 \eqend{.}
\end{equations}
Probably the most striking fact from Eq.~\eqref{eq:coefficients_anomaly} is the vanishing of the coefficient $f$ for all three types of particles considered.

Recently, Bonora et al.\ have claimed that this impasse is broken by Weyl fermions.
In particular, in Refs.~\cite{Bonora:2014qla,Bonora:2015nqa, Bonora:2017gzz} they obtain a non-vanishing imaginary value for $f$; 
in spite of this being consistent with CPT symmetry,
it would in principle endanger the unitarity of the standard model, which is known to contain a net chirality as a consequence of neutrinos.
At the same time, it could offer opportunities to understand some open questions of the early universe, such as the way in which the CP symmetry could have been broken and how one could avoid a Big Bang scenario~\cite{Dolgov,Fabris:1998vq,Hawking:2000bb,CesareSilva:2020ihf}.

From a cohomological point of view, such an anomaly is consistent~\cite{Bonora:1984ic, Bonora:1985cq} and cannot be discarded. Bonora et al.\ have employed two methods in its derivation: a diagrammatic approach~\cite{Bonora:2014qla} and a heat-kernel one~\cite{Bonora:2017gzz}; in the latter,  a chiral component is introduced in the metric, which is then generalized to a metric-axial-tensor. The nuisance is that other groups have failed to derive this non-vanishing $f$. A variety of approaches have all obtained a vanishing $f$, including a Pauli--Villars regularization~\cite{Bastianelli:2016nuf, Bastianelli:2019zrq}, a Hadamard approach~\cite{Frob:2019dgf}, as well as another Feynman diagrammatic method~\cite{Abdallah:2021eii,Abdallah:2022okt}. The first independent support to Bonora's group arrived last year, when a theory of massive fermions (in the massless limit) was considered~\cite{Liu:2022jxz}. In the meanwhile, the discrepancy has been extended to parity-odd gauge contributions to the Weyl anomaly~\cite{Bastianelli:2018osv, Bastianelli:2019fot, Bonora:2020upc, Bastianelli:2022hmu}.

In the following we will delve into this discussion, showing some details of the computation of the trace anomaly for a Weyl fermion employing Feynman diagrams. However, before getting into the specifics, in Sec.~\ref{sec:dirac_fermions} we will provide some basics of Dirac fermions in curved space. Afterwards, in Sec.~\ref{sec:weyl_fermions} we will introduce Weyl fermions and all the ingredients necessary for a perturbative computation of the trace anomaly, including the dimensional regularization of the chiral matrix $\gamma_*$ developed by Breitenlohner and Maison~\cite{Breitenlohner:1977hr}. Concrete results on the trace anomaly for Weyl fermions are given first in Sec.~\ref{sec:trace_anomaly}, while their discussion is left to Sec.~\ref{sec:conclusions}.

\section*{Conventions}
We choose to follow the conventions in Ref.~\cite{Freedman:2011hp}. This includes a mostly plus metric $(-,+,+,\cdots)$, a factor $\mathi$ absorbed in the definition of the adjoint spinor $\bar{\psi} \coloneqq \mathi \psi^* \gamma^0$, the Riemann tensor is defined as $R^\sigma{}_{\mu\rho\nu} \coloneqq \partial_\rho \Gamma^\sigma_{\mu\nu} + \cdots$ and upon contraction gives the Ricci tensor $R_{\mu\nu} \coloneqq R^\rho{}_{\mu\rho\nu}$. We use natural units ($c = \hbar = 1$) and normalize the totally antisymmetric symbol as $\epsilon_{0123} \coloneqq 1$.

\section{Dirac fermions in curved space} \label{sec:dirac_fermions}

Let us recall some notions needed to work with fermions in curved spacetimes. First of all, consider an $n$-dimensional pseudo-Riemannian manifold $\mathcal{M}$ described by the metric $g_{\mu\nu}$. On $\mathcal{M}$, the introduction of Dirac fermions, i.e. quantities transforming under the $(\tfrac{1}{2},0)\oplus(0,\tfrac{1}{2})$ representation of the Lorentz group, is well-established\footnote{We will of course assume that the manifold admits a spinor bundle; a necessary and sufficient condition is that the first and the second Stiefel--Whitney classes of the manifold must be trivial~\cite{DeWitt:2003, Choquet-Bruhat:2000}, i.e., $H^1(M;\mathbb{Z}_2) = 0 = H^2(M;\mathbb{Z}_2)$. A more intuitive notion is that of parallelizable manifolds, which are those that admit globally defined vector fields that are orthonormal; parallelizability is a sufficient condition for the global existence of a Dirac spinor.}~\cite{Toms:2009, DeWitt:2003, Srednicki:2010}. The basic ingredients are the Dirac gamma matrices $\gamma^a$ with anticommutators related to the Minkowski metric $\eta^{ab}$,
\begin{equation}
\left\{ \gamma^a, \gamma^b \right\} = 2 \eta^{ab} \eqend{,}
\end{equation}
and the vielbein $e_\mu{}^a$ (also called non-holonomic connection or, in four dimensions, tetrad), which locally trivializes the tangent bundle of $\mathcal{M}$:
\begin{equation}
g_{\mu\nu} = e_\nu{}^a \eta_{ab} e_\mu{}^b \eqend{.}
\end{equation}
Combining both the gamma matrices and the vielbein, one can construct the so-called curved-space gamma matrices, $\underline{\gamma}^\mu$, defined as\footnote{In the following we will omit the underline whenever the nature of the gamma matrix can be understood from its index.}
\begin{equation}
\underline{\gamma}^\mu \coloneqq g^{\mu\nu} e_\nu{}^a \eta_{ab} \gamma^b \eqend{,}
\end{equation}
which satisfy the anticommutation relation
\begin{equation}
\left\{ \gamma^\mu, \gamma^\nu \right\} = 2 g^{\mu\nu} \eqend{.}
\end{equation}
We can also build higher-order gamma matrices by antisymmetrization of products\footnote{Idempotent symmetrization (antisymmetrization) is indicated by parentheses (brackets); for example, $v^{[a} w^{b]} = \tfrac{1}{2} \left( v^a w^b - v^b w^a \right)$.}~\cite{Kennedy:1981kp}, $\gamma^{\mu_1 \cdots \mu_k} \coloneqq \gamma^{[\mu_1} \cdots \gamma^{\mu_k]}$.

To construct an action for a free fermion, one needs a covariant derivative $\nabla_\mu$ compatible with the fermionic structure; this is accomplished by the use of the spin connection $\omega_\mu{}^{ab}$, which upon contraction with two vielbeins can be written as\footnote{We assume the torsion-free condition.}~\cite{Freedman:2011hp}
\begin{equation}
\label{eq:spin_connection}
\omega_{\mu\rho\sigma} = \omega_{\mu[\rho\sigma]} = \eta_{ab} \left( e_\sigma{}^a \partial_{[\mu} e_{\rho]}{}^b - e_\rho{}^a \partial_{[\mu} e_{\sigma]}{}^b + e_\mu{}^a \partial_{[\sigma} e_{\rho]}{}^b \right) \eqend{.}
\end{equation}
In terms of this spin connection and the second-order gamma matrices, the covariant derivative has the compact expression
\begin{equation}
\label{eq:covariant_derivative}
\nabla_\mu \psi \coloneqq \partial_\mu \psi + \frac{1}{4} \omega_{\mu\nu\sigma} \gamma^{\nu\sigma} \psi \eqend{,}
\end{equation}
which leads to the following action for a massless fermion:
\begin{equation}
\label{eq:action_dirac}
S_D = \int \mathcal{L}_D \total^n x \coloneqq - \int \bar{\psi} \gamma^\mu \overleftrightarrow{\nabla}_\mu \psi \sqrt{-g} \total^n x \eqend{.}
\end{equation}
We are restricting the discussion to massless fermions in order to obtain a Weyl invariant action, as we will shortly see.
Notice that in order to work with a Hermitian operator,
in Eq.~\eqref{eq:action_dirac} we have introduced an operator involving right and left derivatives, condensed in the symbol $\overleftrightarrow{\nabla}_\mu \coloneqq \nabla_\mu - \overleftarrow{\nabla}_\mu$; in particular, the derivative to the left $\overleftarrow{\nabla}_\mu$ acts in the following way:
\begin{equation}
\bar{\psi} \overleftarrow{\nabla}_\mu \coloneqq \partial_\mu \bar{\psi} - \frac{1}{4} \bar\psi\, \omega_{\mu\nu\sigma} \gamma^{\nu\sigma} \eqend{.}
\end{equation}

The free Dirac action in Eq.~\eqref{eq:action_dirac} possesses several symmetries. Its invariance under diffeomorphisms and under Lorentz transformations is evident by construction. More interestingly, it is also invariant under $n$-dimensional Weyl transformations
\begin{equations}
\psi(x) &\to \psi'(x) = \mathe^{-\frac{(n-1)}{2}\,\sigma(x)} \psi(x) \eqend{,} \\
e_\mu{}^a &\to e'_\mu{}^a = \mathe^{\sigma(x)} e_\mu{}^a \eqend{.}
\end{equations}
The Weyl invariance of the action can be straightforwardly checked by considering the transformation induced in the spin connection [using Eq.~\eqref{eq:spin_connection} as definition] and the properties of the gamma matrices, to wit
\begin{equations}
\omega_{\mu\rho\nu} &\to \omega'_{\mu\rho\nu} = \omega_{\mu\rho\nu} + g_{\mu[\rho} \partial_{\nu]} \sigma(x) \eqend{,} \\
\gamma_\mu \gamma^{\mu\nu} &= 2 (n-1) \gamma^\nu \eqend{.}
\end{equations}
An immediate consequence of this symmetry is the tracelessness of the on-shell EM tensor in the classical theory. Indeed, the EM tensor is computed as a variation of the action with respect to the vielbein\footnote{A generic variation of the vielbein can be decomposed into symmetric and antisymmetric parts~\cite{Forger:2003ut}; the symmetric part is determined by the variation of the metric, while the antisymmetric part is a local Lorentz transformation. Since the action is invariant under Lorentz transformations, only the metric variation contributes;
thus, the variation can be understood in terms of the metric, as for bosonic fields.}:
\begin{equation}
T^{\mu\nu} = \frac{1}{\sqrt{-g}} \frac{\delta S_D}{\delta e_{\mu a}} e^\nu{}_a \eqend{,}
\end{equation}
so that under a Weyl transformation the chain rule can be used to obtain
\begin{splitequation}
0 &= \frac{\delta S_D}{\delta \sigma} = \frac{\delta S_D}{\delta e_{\mu a}} \frac{\delta e_{\mu a}}{\delta \sigma} + \frac{\delta S_D}{\delta \psi} \frac{\delta \psi}{\delta \sigma} + \frac{\delta S_D}{\delta \bar{\psi}} \frac{\delta \bar{\psi}}{\delta \sigma} \\
&= \sqrt{-g} \, T^\mu{}_\mu \quad (\text{on-shell}) \eqend{.}
\end{splitequation}

It can also be proved that the EM tensor is both symmetric and conserved, respectively due to the invariance under Lorentz transformations and diffeomorphisms. While its symmetry can be achieved off-shell, conservation only holds on-shell. To verify these facts, consider the explicit expression for the EM tensor~\cite{Forger:2003ut}:
\begin{equation}
\label{eq:EMtensor_dirac}
T^{\mu\nu}_D = \frac{1}{2} \bar{\psi} \gamma^{(\mu} \overleftrightarrow{\nabla}^{\nu)} \psi + \frac{1}{2} g^{\mu\nu} \bar{\psi} \gamma^\rho \overleftrightarrow{\nabla}_\rho \psi \eqend{.}
\end{equation}
Its symmetry is evident, while its conservation follows easily once one realizes that the curvature in the fermionic bundle is given by~\cite{Freedman:2011hp}
\begin{equation}
\left[ \nabla_\mu, \nabla_\nu \right] \psi = \tfrac{1}{4} R_{\mu\nu\rho\sigma} \gamma^{\rho\sigma} \psi \eqend{.}
\end{equation}
Note that the second term in Eq.~\eqref{eq:EMtensor_dirac} is usually dismissed, being null on-shell; taking into account that the computation of anomalies is full of subtleties, we will keep it in the following.

As discussed above in Sec.~\ref{sec:introduction}, it is well-known that the trace of the EM tensor develops an anomaly at the quantum level~\cite{Capper:1974ic, Capper:1975ig}, whose coefficients for a Dirac fermion are~\cite{DeWitt:2003,Godazgar:2018boc}:
\begin{equation}
\label{eq:coefficients_Dirac}
w_D = \frac{18}{360 (4 \pi)^2} \eqend{,} \quad b_D = - \frac{11}{360 (4 \pi)^2} \eqend{,} \quad c_D = \frac{12}{360 (4 \pi)^2} \eqend{,} \quad f = 0 \eqend{.}
\end{equation}
These coefficients have been computed in several different ways, employing techniques such as $\zeta$ function regularization~\cite{DeWitt:2003} and Feynman diagrams. The latter involves the famous triangle diagram, this time containing three insertions of $T^{\mu\nu}_D$. To avoid repetition in the calculations, the triangle diagram will be introduced first in Sec.~\ref{sec:trace_anomaly} for a Weyl fermion; the results for a Dirac fermion can be obtained from these expressions by simply dropping the chiral projectors.

\section{Weyl fermions in curved space}
\label{sec:weyl_fermions}

In this article, we are not interested in Dirac fermions but in right-handed Weyl fermions $\psi_R$, which are defined as spinors in the $(0,\frac{1}{2})$ representation of the Lorentz group; equivalently, one may obtain a right-handed Weyl fermion acting with a chiral projector on a Dirac fermion~$\psi$,
\begin{equation}
\label{eq:def_weyl_fermion}
\psi_R \coloneqq \mathcal{P}_+ \psi \coloneqq \frac{1}{2} \left( \1 + \gamma_* \right) \psi \eqend{,}
\end{equation}
where $\gamma_*$ (also called $\gamma_{n+1}$ in the literature) is the chiral gamma matrix satisfying 
\begin{equation}
\label{eq:gammastar_properties}
\gamma_*^2 = \1 \eqend{,} \quad \gamma_*^\dagger = \gamma_* \eqend{,} \quad \left\{ \gamma^\mu, \gamma_* \right\} = 0 \eqend{.}
\end{equation}
Notice that for a right-handed Weyl fermion, employing these properties of $\gamma_*$, the adjoint spinor turns out to be left-handed, meaning that it transforms with a projector similar to that in Eq.~\eqref{eq:def_weyl_fermion} but with a reversed sign in the chiral term:
\begin{equation}
\overline{\psi_R} = \bar{\psi} \mathcal{P}_- \coloneqq \frac{1}{2} \bar{\psi} \left( \1 - \gamma_* \right) \eqend{.}
\end{equation}
Indeed, this is natural recalling that, upon a parity transformation, the $\gamma_*$ transforms to $(- \gamma_*)$. One can also verify that the operators $\mathcal{P}_\pm$ are projectors (by showing that they are idempotent $\mathcal{P}_\pm^2 = \mathcal{P}_\pm$) and that they are orthogonal $\mathcal{P}_\pm \mathcal{P}_\mp = 0$, both of which follow from the fact that $\gamma_*$ squares to the identity.

Now suppose that we want to write down an action for our right-handed Weyl fermion. One first trial could  be to introduce projectors in the action of a Dirac fermion; in spite of the simplicity of this idea, some difficulties would be met. 

On the one hand, a Dirac mass term is not allowed, i.e., Weyl fermions are massless; indeed such a term would not be Lorentz invariant. An alternative way to confirm this fact is by explicitly introducing the projectors that correspond to the fermion and its adjoint: $\overline{\psi_R} \psi_R = \bar{\psi} \mathcal{P}_+ \mathcal{P}_- \psi = 0$. 
In any case, this is not a real obstacle, since a mass term would have broken Weyl invariance.

On the other hand, it is permitted to build a kinetic term involving the covariant derivative, namely
\begin{equation}
\label{eq:action}
S = - \int \bar{\psi} \mathcal{P}_- \gamma^\mu \nabla_\mu \left( \mathcal{P}_+ \psi \right) \sqrt{-g} \total^n x \eqend{.}
\end{equation}
However, this definition leads to an inconvenience at the moment of defining a quantum theory: given the presence of the projector in this kinetic term, the relevant operator turns out to be non-invertible in the space of Dirac fermions, or, to be more explicit, as an operator it maps left-handed fermions to right-handed fermions. To circumvent this problem, we will not project out the left-handed fermion; instead, similar to the proposal in Ref.~\cite{AlvarezGaume:1983cs}, we will keep it but decoupled from gravity. Let us describe how this could be done in the following paragraphs. 

Since we are interested in a perturbative computation, it is natural to introduce the quantity $h_{\mu\nu}$ which measures the deviation of the metric from the flat metric in Minkowski space:
\begin{equation}
\kappa h_{\mu\nu} \coloneqq g_{\mu\nu} - \eta_{\mu\nu} \eqend{.}
\end{equation}
In the following, $\kappa$ will be used as the parameter in a perturbative expansion (if we were quantizing the theory, it would be linked to the Planck mass). In particular, all the geometric quantities may be expanded in powers of $\kappa$; as we will see, for the determination of the trace anomaly it would suffice to work up to second order in $\kappa$, such that we have
\begin{equations}[{eq:geometric_expansion}]
e^\mu{}_a &= e^{(0)}{}^\rho{}_a \left( \eta^\mu{}_\rho - \frac{1}{2} \kappa h^\mu{}_\rho + \frac{3}{8} \kappa^2 h^{\mu\sigma} h_{\sigma\rho} \right) + \bigo{\kappa^3} \eqend{,} \\
e_\mu{}^a &= e^{(0)}{}^a{}_\rho \left( \eta^{\mu\rho} + \frac{1}{2} \kappa h^{\mu\rho} - \frac{1}{8} \kappa^2 h^{\mu\sigma} h_\sigma{}^\rho \right) + \bigo{\kappa^3} \eqend{,} \\
g^{\mu\nu} &= \eta^{\mu\nu} - \kappa h^{\mu\nu} + \kappa^2 h^{\mu\rho} h_\rho{}^\nu + \bigo{\kappa^2} \eqend{,} \\
\sqrt{-g} &= 1 + \frac{\kappa}{2} h^\rho{}_\rho + \frac{\kappa}{8} \left( h^\rho{}_\rho h^\sigma{}_\sigma - 2 h^{\rho\sigma} h_{\rho\sigma} \right) + \bigo{\kappa^3} \eqend{,}
\end{equations}
where $e^{(0)}{}^\rho{}_a = \delta^\rho_a$ denotes the vielbein in Minkowski space. Notice that in writing the expansions in Eq.~\eqref{eq:geometric_expansion} we have used the following conventions: we have employed the symmetric gauge for the vielbein~\cite{Woodard:1984sj} and the indices in the RHS have been raised with a flat metric (in all cases). Even if the former is simply a choice and thus irrelevant for the computations, the latter is an abuse of notation and crucial for rewriting the final results in a covariant way.

Armed with these transformations, one can readily expand any further relevant object in the theory; the notation of such an expansion for an object $X$ will be 
\begin{equation}
X \coloneqq X_{(0)} + \kappa X_{(1)} + \kappa^2 X_{(2)} + \bigo{\kappa^3} \eqend{.}
\end{equation}
Coming back to our idea of introducing a left-handed Weyl fermion decoupled from gravity, this can be done by expanding the action in powers of $\kappa$ and introducing projectors in all the interaction terms, such that only the right-hand part of the Dirac fermion couples to gravity. Proceeding in this way, the action which we are going to consider has the expansion coefficients
\begin{equations}[{eq:action_expansion}]
S_{(0)} &\coloneqq - \frac{1}{2} \int\left[ \Psi^{\mu}{}_{\mu} +\left( \Psi^{\mu}{}_{\mu}\big\vert_{+\leftrightarrow -}\right)\right] \total^n x \eqend{,} \label{eq:actionexpansion_s0} \\
S_{(1)} &\coloneqq \frac{1}{4} \int \left( h_{\alpha\beta} \Psi^{\alpha\beta} - h^{\mu}{}_{\mu} \Psi^{\nu}{}_{\nu} \right) \total^n x \eqend{,} \label{eq:actionexpansion_s1} \\
S_{(2)} &\coloneqq \frac{1}{16} \int \left[ \left( 2 h^{\mu}{}_{\mu} h_{\alpha\beta} - 3 h_\alpha{}^\delta h_{\beta\delta} \right) \Psi^{\alpha\beta} + \left( 2 h_{\alpha\beta} h^{\alpha\beta} - \left(h^{\mu}{}_{\mu}\right)^2 \right) \Psi^{\nu}{}_{\nu} + h_\alpha{}^\delta \partial_\gamma h_{\beta\delta} J^{\alpha\beta\gamma} \right] \total^n x \eqend{,} \label{eq:actionexpansion_s2}
\end{equations}
where in the first line $(+\leftrightarrow-)$ denotes the exchange of the chiral projectors $\mathcal{P}_{+}\leftrightarrow \mathcal{P}_{-}$ to obtain the free action for a left-handed fermion.
Moreover, we have defined the basic structures\footnote{Recall that at this point we are contracting indices with a flat metric; additionally, the gamma matrices in the expressions from now on will always be flat gamma matrices.}
\begin{equations}[eq:structures]
\Psi^{\mu\nu} &\coloneqq \bar\psi \mathcal{P}_- \gamma^\mu \mathcal{P}_+ \partial^\nu \psi - \partial^\nu \bar\psi \mathcal{P}_- \gamma^\mu \mathcal{P}_+ \psi \eqend{,} \\
J^{\mu\nu\rho} &\coloneqq \bar\psi \mathcal{P}_- \gamma^{\mu\nu\rho} \mathcal{P}_+ \psi \eqend{.}
\end{equations}
Physically, we are incorporating the information of the chirality in the interaction sector, which at this order corresponds to cubic and quartic interactions between the right-handed fermion and gravity; as a consequence, one would expect that physical quantities which depend on the geometry should not be affected by the presence of the left-handed companion. Even if this is also the path followed by Refs.~\cite{Bonora:2014qla,Bonora:2015nqa}, there is one difference: in our case, we have included a projector for each single fermionic operator, which we are not going to remove employing the usual four-dimensional identities. This is a subtle point, related to the fact that we are going to employ dimensional regularization; a more detailed account on this will be given in Sec.~\ref{sec:BMprescription}. By now we can say that, in spite of giving the same expression when naively setting $n = 4$, one would expect that the results differ by some evanescent terms in the limit when $n$ tends to four~\cite{Bollini:1973wu}. Therefore, it is more prudent to keep all these projectors and only use four-dimensional identities to remove them at the very end of the computation.

\subsection{Energy-momentum tensor}

Before discussing regularization issues, let us focus on the most relevant object for this article: the EM tensor $T^{\mu\nu}$. In any quantum field theory, it is promoted to be an operator; its vacuum expectation value (VEV), for a theory with action $S$ and in a curved spacetime with background metric $g$, will be written as $\langle T^{\mu\nu} \rangle_g$. Thanks to the Gell-Mann--Low formula, the latter can be expressed in terms of VEVs in a theory with action $S_{(0)}$ and in flat spacetime, which we will simply write as $\langle \cdot \rangle$:
\begin{equation}
\expect{ T^{\mu\nu}(x) }_g = \frac{\expect{ T^{\mu\nu}(x) \exp\left[ \mathi \left( \kappa S_{(1)} + \kappa^2 S_{(2)} \right) \right]}}{\expect{ \exp\left[ \mathi \left( \kappa S_{(1)} + \kappa^2 S_{(2)} \right) \right]}} + \bigo{\kappa^3} \eqend{.}
\label{eq:vev_em_tensor}
\end{equation}
Performing an expansion in powers of $\kappa$ and omitting terms that vanish in dimensional regularization\footnote{Recall that in a massless theory and using dimensional regularization, the vacuum expectation value of any operator which depends on just one position vanishes~\cite{Leibbrandt:1975dj}; for example $\expect{ T^{\mu\nu}_{(2)}(x) }$ and $\expect{ S_{(2)}(x) }$ both vanish.} we are lead to the following coefficients in the perturbative expansion of the VEV of the EM tensor:
\begin{equations}[eq:coefficients_VEV]
\expect{ T^{\mu\nu}(x) }_{(1)} &= \mathi \expect{ S_{(1)} T^{\mu\nu}_{(0)}(x) } \eqend{,} \\
\expect{ T^{\mu\nu}(x) }_{(2)} &= - \frac{1}{2} \expect{ S_{(1)}^2 T^{\mu\nu}_{(0)}(x) } + \mathi \expect{ S_{(2)} T^{\mu\nu}_{(0)}(x) } + \mathi \expect{ S_{(1)} T^{\mu\nu}_{(1)}(x) } \eqend{.}
\end{equations}
To further simplify, notice that to obtain the trace anomaly for Weyl fermions the relevant part of the EM momentum operator is only the right-handed one; this means that, instead of using the full expression in Eq.~\eqref{eq:EMtensor_dirac}, we can safely work with the contribution 
\begin{equation}
\label{eq:EMtensor_weyl}
T^{\mu\nu} = \frac{1}{2} \bar{\psi} \mathcal{P}_- \gamma^{(\mu} \mathcal{P}_+ \overleftrightarrow{\nabla}^{\nu)} \psi + \frac{1}{2} g^{\mu\nu} \bar{\psi} \mathcal{P}_- \gamma^\rho \mathcal{P}_+ \overleftrightarrow{\nabla}_\rho  \psi \eqend{.}
\end{equation}
Taking into account the expansions in Eq.~\eqref{eq:geometric_expansion}, it is straightforward to obtain the only missing ingredient to compute the VEVs in  Eq.~\eqref{eq:coefficients_VEV}, i.e., the following expansion for the EM operator:
\begin{equations}[eq:emtensorexpansion]
T^{\mu\nu}_{(0)} &= \frac{1}{2} \left[ \Psi^{(\mu\nu)} - \Psi^{\rho}{}_{\rho} \eta^{\mu\nu} \right] \eqend{,} \label{eq:emtensorexpansion_0} \\
T^{\mu\nu}_{(1)} &= \frac{1}{4} \left[ 2 \Psi^{\rho}{}_{\rho} h^{\mu\nu} + h_{\alpha\beta} \Psi^{\alpha\beta} \eta^{\mu\nu} - 2 h^{\alpha(\mu} \Psi^{\nu)}{}_\alpha - h^{\alpha(\mu} \Psi_\alpha{}^{\nu)} - J^{\alpha\beta(\mu} \partial_\alpha h^{\nu)}{}_\beta \right] \eqend{,} \label{eq:emtensorexpansion_1} \\
\begin{split}
T^{\mu\nu}_{(2)} &= \frac{1}{16} \Big[ 4 h^{\rho(\mu} h^{\nu)\sigma} - 4 h_{\rho\sigma} h^{\mu\nu} + 8 h^{\alpha(\mu} \left( \delta^{\nu)}_\rho h_{\alpha\sigma} - h^{\nu)}{}_\alpha \eta_{\rho\sigma} \right) \\
&\qquad+ 3 h_{\alpha\rho} \left( h^{\alpha(\mu} \delta^{\nu)}_\sigma - h_\sigma{}^\alpha \eta^{\mu\nu} \right) \Big] \Psi^{\rho\sigma} + \frac{1}{16} \Big[ \left( 4 h^{\alpha(\mu} \eta^{\nu)\rho} - \eta^{\mu\nu} h^{\alpha\rho} \right) \partial^\sigma h_\alpha{}^\tau \\
&\hspace{3em}+ 2 h^{\mu\rho} \partial^\sigma h^{\nu\tau} - h^{\alpha\sigma} \eta^{\rho(\mu} \left( \partial^{\nu)} h_\alpha{}^\tau - 2 \partial_\alpha h^{\nu)}{}_\tau + 2 \partial_\tau h^{\nu)}{}_\alpha \right) \Big] J_{\rho\sigma\tau} \eqend{.} \label{eq:emtensorexpansion_2}
\end{split}
\end{equations}

\begin{figure}
\centering
\begin{subfigure}[h]{0.3\textwidth}
\centering
\begin{tikzpicture}[scale=1.1,every node/.style={transform shape}]
 \begin{feynman}
  \vertex  (T) ;
  \vertex [right= of T] (f1);
  \vertex [right= of f1] (f2);
    \vertex [right= of f2] (h1) {\(h^{\alpha\beta}\)};
 
 \diagram*[]{
 (T) -- [gluon, momentum=\(q_1\), insertion={[size=5pt]0}] (f1) -- [fermion, half left] (f2) -- [fermion, half left] (f1),
(f2) -- [gluon, momentum=\(q_2\)] (h1),
 };
 \end{feynman}
\end{tikzpicture}
\caption{}
\label{fig:orderk}
\end{subfigure}

\begin{subfigure}[h]{0.48\textwidth}
\centering
\begin{tikzpicture}[scale=1.1,every node/.style={transform shape}]
 \begin{feynman}
  \vertex (T) ;
  \vertex [right= of T] (f1) ;
  \vertex [right= of f1] (f2);
 \vertex [above right= of f2] (h1) {$h^{\alpha\beta}$};
  \vertex [below right= of f2] (h2) {$h^{\rho\sigma}$};
 
 \diagram*[]{
 (T)-- [gluon, insertion={[size=5pt]0}, momentum=\(q_1\)] (f1) -- [fermion, half left] (f2) -- [fermion, half left] (f1),
(f2) -- [gluon, momentum=\(q_2\)] (h1),
(f2) -- [gluon, momentum'=\(q_3\)] (h2),
 };
 \end{feynman}
\end{tikzpicture}
\caption{}
\label{fig:orderk2_bubble}
\end{subfigure}
\begin{subfigure}[h]{0.48\textwidth}
\centering
\begin{tikzpicture}[scale=1.1,every node/.style={transform shape}]
 \begin{feynman}
  \vertex (T) ;
  \vertex [right= of T] (f1);
  \vertex [above right= of f1] (f2);
  \vertex [below right= of f1] (f3);
  \vertex [right= of f2] (h1) {\(h^{\alpha\beta}\)};
  \vertex [right= of f3] (h2) {\(h^{\rho\sigma}\)};
 
 \diagram*[]{
 (T) -- [gluon, momentum=\(q_1\), insertion={[size=5pt]0}] (f1) -- [fermion] (f2) -- [fermion] (f3) -- [fermion] (f1),
(f2) -- [gluon, momentum=\(q_2\)] (h1),
(f3) -- [gluon, momentum=\(q_3\)] (h2),
 };
 \end{feynman}
\end{tikzpicture}
\caption{}
\label{fig:orderk2_triangle}
\end{subfigure}

\caption{Prototype of the Feynman diagrams involved in the computation: the bubble in Fig.~\ref{fig:orderk} corresponds to order $\kappa$, while the bubble and triangle (respectively in Figs.~\ref{fig:orderk2_bubble} and \ref{fig:orderk2_triangle}) contribute at order $\kappa^2$. 
As usual, solid lines represent fermions and curly lines represent external metric factors ($h^{\mu\nu}$);
a crossed insertion denotes an amputated external $h^{\mu\nu}$ factor.}
\label{fig:generic_diagrams}
\end{figure}
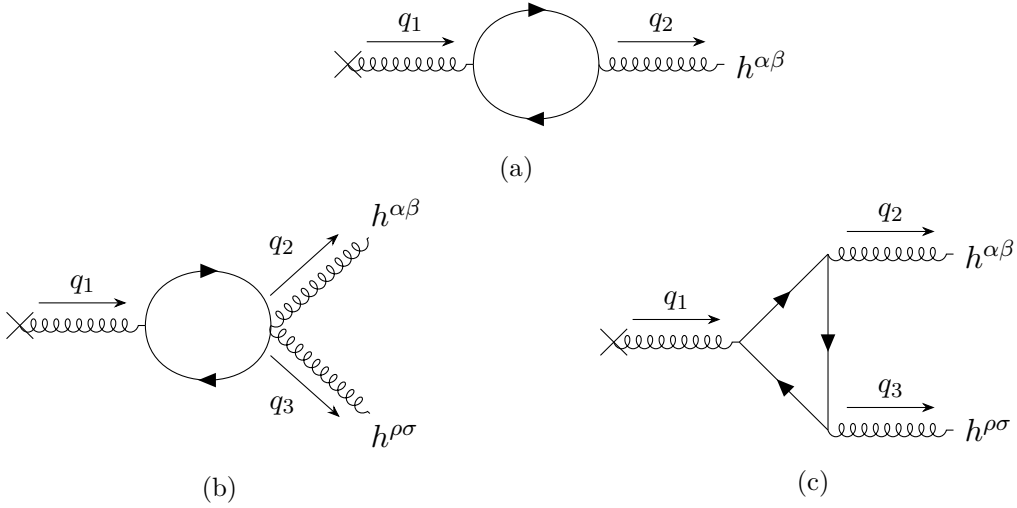

As customarily, joining the expansions in Eqs.~\eqref{eq:action_expansion} and \eqref{eq:emtensorexpansion}, the corresponding Feynman diagrams might be defined; generically, they will be of the form depicted in Fig.~\ref{fig:generic_diagrams}, where there will be a few types of different vertices. In the computation of their contributions, divergences will appear and we will have to resort to a regularizing prescription, which we choose to be dimensional regularization; 
its implementation is not straightforward, since our chiral theory involves the further difficulty of handling the $\gamma_*$ matrix~(see Sec.~\ref{sec:BMprescription}). 
Regarding the notation in this process, in the VEVs we will use the superscripts ``reg'', ``div'' and ``ren'', to respectively denote the quantities computed in $n$ dimensions, the divergent terms in the limit $n \to 4$ and the results obtained after  subtracting the divergences.

Once the computation of the VEVs has been performed, one still has to compute the trace $\mathcal{T} \coloneqq g_{\mu\nu} \expect{ T^{\mu\nu} }_g$. Its first coefficients in the perturbative expansion can be written as
\begin{equations}[eq:trace_definitions]
\mathcal{T}_{(1)}(x) &= \eta_{\mu\nu} \expect{ T^{\mu\nu}(x) }_{(1)} \eqend{,} \label{eq:tr_orderh} \\
\mathcal{T}_{(2)}(x) &= \eta_{\mu\nu} \expect{ T^{\mu\nu}(x) }_{(2)} + h_{\mu\nu} \expect{ T^{\mu\nu}(x) }_{(1)} \eqend{.} \label{eq:tr_orderh2}
\end{equations}

Finally, there is one further check to be done. As a consequence of invariance under diffeomorphisms, the classical theory possesses an EM tensor that is conserved; this is a symmetry that we would like to preserve also at the quantum level, and, as a consequence of the triviality of the corresponding cohomology, in $n = 4$ one is always able to do so (possibly after adding some finite counterterms). Of course, the addition of these counterterms might also change the VEV of the trace in Eq.~\eqref{eq:vev_em_tensor}. Having this in mind, we write down the relevant expressions for the divergence of the VEV of the EM operator; up to second order in $\kappa$, we obtain
\begin{equations}[eq:div_definitions]
\mathcal{D}^\nu_{(1)}(x) &= \partial_\mu \expect{ T^{\mu\nu}(x) }_{(1)} \eqend{,} \label{eq:div_orderh} \\
\mathcal{D}^\nu_{(2)}(x) &= \partial_\mu \expect{ T^{\mu\nu}(x) }_{(2)} + \frac{1}{2} \left( 2 \partial_\rho h^\nu{}_\mu - \partial^\nu h_{\rho\mu} + \delta^\nu_\mu \partial_\rho h \right) \expect{ T^{\mu\rho}(x) }_{(1)} \eqend{.} \label{eq:div_orderh2}
\end{equations}

\subsection{Breitenlohner--Maison prescription}
\label{sec:BMprescription}

Let us finally turn our attention to the regularization procedure. The first choice, due to its versatility and simplicity in the computations involved, is dimensional regularization~\cite{Bollini:1972ui,tHooft:1972tcz,Leibbrandt:1975dj}. 
As we will show, the fact that we are dealing with a chiral theory will cause some headaches, inasmuch as one would like to keep the usual properties of $\gamma_*$. 
In fact, notice that the object defined in $n = 4$ as
\begin{equation}
\gamma_*^{n=4} \coloneqq - \frac{\mathi}{4!} \epsilon_{\mu\nu\rho\sigma} \gamma^\mu \gamma^\nu \gamma^\rho \gamma^\sigma \eqend{,}
\end{equation}
satisfies the well-known properties 
\begin{equations}[eq:properties_g5]
\tr\left( \gamma_*^{n=4} \gamma_\mu \gamma_\nu \gamma_\rho \gamma_\sigma \right) &= - \mathi \, \epsilon_{\mu\nu\rho\sigma} \tr \1 \eqend{,} \label{eq:g5fourg} \\
\left\{ \gamma_\mu, \gamma_*^{n=4} \right\} &= 0 \eqend{,}\label{eq:g5ganticommute}
\\
\left(\gamma_*^{n=4}\right)^2 &= \1 \eqend{.} \label{eq:g5square}
\end{equations}
If they were valid for arbitrary $n$, we would be led into a contradiction. To see this, first notice that we could show that $n \tr \gamma_* = 0$ in the following way:
\begin{equation}
n \tr \gamma_* = \tr\left( \gamma_* \gamma^\alpha \gamma_\alpha \right) = - \tr\left( \gamma^\alpha \gamma_* \gamma_\alpha \right) = - \tr\left( \gamma_* \gamma_\alpha \gamma^\alpha \right) = - n \tr \gamma_* \eqend{.}
\end{equation}
In this expression we have used the anticommutator in Eq.~\eqref{eq:g5ganticommute} to anticommute the $\gamma_*$ with another $\gamma$ matrix and afterwards the cyclicity of the trace. In a subsequent step, we could introduce a pair of gamma matrices inside the trace; \emph{mutatis mutandis}, we would obtain a series of equalities
\begin{splitequation}
n \tr\left( \gamma_* \gamma_\mu \gamma_\nu \right) &= \tr\left( \gamma_* \gamma^\alpha \gamma_\alpha \gamma_\mu \gamma_\nu \right) = - \tr\left( \gamma_* \gamma_\alpha \gamma_\mu \gamma_\nu \gamma^\alpha \right) \\
&= \ldots = - 2 \tr\left( \gamma_* \gamma_\nu \gamma_\mu \right) + 2 \tr\left( \gamma_* \gamma_\mu \gamma_\nu \right) - n \tr\left( \gamma_* \gamma_\mu \gamma_\nu \right) \\
&= - 4 g_{\mu\nu} \tr\gamma_* - (n-4) \tr\left( \gamma_* \gamma_\mu \gamma_\nu \right)\eqend{,}
\end{splitequation}
which would imply that $n (n-2) \tr\left( \gamma_* \gamma_\mu \gamma_\nu \right) = 0$. Adding still one further pair of gamma matrices inside the trace, one would be lead to the conclusion that 
\begin{equation}
\label{eq:gammastar_trace_ndim}
n (n-2) (n-4) \tr\left( \gamma_* \gamma_\mu \gamma_\nu \gamma_\rho \gamma_\sigma \right) = 0 \eqend{,}
\end{equation}
which is in clear contradiction with Eq.~\eqref{eq:g5fourg} if $n \neq 4$.

To solve this conundrum we are forced to renounce (at least) one of the properties that were employed in this series of equalities: either we have to abandon the cyclicity of the trace (embedding for example the Clifford algebra in an infinite-dimensional one~\cite{Korner:1991sx,Kreimer:1989ke}), give up the anticommutation property of $\gamma_*$ with other gamma matrices in a covariant manner~\cite{Thompson:1985uv} (giving rise to heavier expressions), or try a prescription that breaks the covariance in the anticommutators $\left\{ \gamma_*,\gamma^\mu \right\}$~\cite{tHooft:1972tcz,Breitenlohner:1977hr}.

For the sake of simplicity, in the following  we will employ the latter, to which we will refer as the Breitenlohner--Maison scheme. Its application in higher-loops contributions to gauge theories has recently been studied in Ref.~\cite{Belusca-Maito:2021lnk}. As a consequence of the break of covariance, one would expect the appearance of spurious non-covariant terms in the calculation of the VEVs; nevertheless, we will be able to remove them by the introduction of finite non-covariant counterterms, as previously done in Refs.~\cite{Bonneau:1980ya,Martin:1999cc,SanchezRuiz:2002xc,Belusca-Maito:2020ala}.

In the Breitenlohner--Maison scheme, the usual Clifford algebra in $n$ dimensions is defined:
\begin{equation}
\left\{ \gamma_\mu, \gamma_\nu \right\} = 2 \eta_{\mu\nu} \1 \eqend{.}
\end{equation}
Subsequently, the $n$-dimensional flat space is split into two subspaces, whose dimensions are $4$ and $(n-4)$. This implies a corresponding splitting in other quantities, such as the metric, the gamma matrices and the momenta. To render this more explicit, we will denote with a bar (respectively a hat) a quantity that has nonvanishing components only in the four-dimensional [in the $(n-4)$-dimensional] space. In this way, we can write
\begin{equation}
\eta_{\mu\nu} = \bar{\eta}_{\mu\nu} + \hat{\eta}_{\mu\nu} \eqend{,} \quad \gamma_\mu = \bar{\gamma}_\mu + \hat{\gamma}_\mu \eqend{,} \quad p_\mu = \bar{p}_\mu + \hat{p}_\mu \eqend{,} \quad \ldots\eqend{,} 
\end{equation}
as well as contractions involving different objects, e.g.
\begin{splitequation}
\eta_\mu{}^\nu \hat{\eta}_{\nu\rho} &= \hat{\eta}_{\mu\nu} \hat{\eta}^\nu{}_\rho = \hat{\eta}_{\mu\rho}\eqend{,}\quad 
\bar{\eta}^{\mu\nu} \hat{\eta}_{\nu\rho} = 0 \eqend{,} \quad \hat{\eta}_\mu{}^\nu p_\nu = \hat{p}_\mu \eqend{,} \quad \bar{\eta}_\mu{}^\nu \gamma_\nu = \bar{\gamma}_\mu \eqend{.}
\end{splitequation} 
The crucial point is the definition of the chiral matrix in terms of the four-dimensional Levi--Civita symbol $\epsilon_{\mu\nu\rho\sigma} = \bar{\epsilon}_{\mu\nu\rho\sigma}$:
\begin{equation}
\label{eq:gamma_star_def}
\gamma_* \coloneqq - \frac{\mathi}{4!} \epsilon_{\mu\nu\rho\sigma} \gamma^\mu \gamma^\nu \gamma^\rho \gamma^\sigma = - \frac{\mathi}{4!} \bar{\epsilon}_{\mu\nu\rho\sigma} \bar{\gamma}^\mu \bar{\gamma}^\nu \bar{\gamma}^\rho \bar{\gamma}^\sigma \eqend{.}
\end{equation}
Since this is now a purely four-dimensional object even though we are working in $n$ dimensions, it enables us to avoid inconsistencies like the one in Eq.~\eqref{eq:gammastar_trace_ndim}.

Building on these definitions, we can derive several properties that are of help in the necessary computations. Instead of Eq.~\eqref{eq:properties_g5}, we now have the following properties~\cite{Breitenlohner:1977hr}:
\begin{equations}
\tr\left( \gamma_* \gamma_\mu \gamma_\nu \gamma_\rho \gamma_\sigma \right) &= - \mathi \, \bar{\epsilon}_{\mu\nu\rho\sigma} \tr \1 \eqend{,} \label{eq:g5fourg_bm} \\
\left\{ \gamma_\mu, \gamma_* \right\} &= \left\{ \hat\gamma_\mu, \gamma_* \right\} = 2 \hat\gamma_\mu \gamma_* \eqend{,} \\
\left[ \gamma_\mu, \gamma_* \right] &= \left[ \bar{\gamma}_\mu, \gamma_* \right] = 2 \bar\gamma_\mu \gamma_* \eqend{,} \\
\gamma_*^2 &= \1 \eqend{.}
\end{equations}
One formula that will be particularly helpful in the computation of the trace anomaly is 
\begin{equation}
\mathcal{P}_\pm \gamma^\mu \mathcal{P}_\mp = \bar\gamma^\mu \mathcal{P}_\mp \eqend{.} \label{eq:gamma_sandwich_projector}
\end{equation}
Note that in strictly $n = 4$ dimensions this identity allows us to remove all chiral projectors except one, which is the path followed by Refs.~\cite{Bonora:2014qla,Bonora:2015nqa}. However, as we have remarked before in Sec.~\ref{sec:weyl_fermions}  and see now in detail, in $n$ dimensions a naive generalization of the identity does not hold, since on the right-hand side we have a four-dimensional $\gamma$ matrix. To be more precise, the difference between the naive generalization of the four-dimensional identity and the true identity~\eqref{eq:gamma_sandwich_projector} is an evanescent term~\cite{Bollini:1973wu} involving $\hat{\gamma}^\mu$, the $(n-4)$-dimensional part of the $\gamma$ matrix. We insist on the fact that in dimensional regularization one has to work with a consistent prescription for the chiral matrix $\gamma_*$, which the Breitenlohner--Maison scheme furnishes, and one cannot use the naive generalization of four-dimensional identities before regularization.

\section{Trace anomaly}
\label{sec:trace_anomaly}

\subsection{Complete result up to second order in $\kappa$}

Although working out a few of the Feynman diagrams involved in the computation of the trace anomaly is not a problem, the full calculation is rather cumbersome. We are not going to show the lengthy intermediate steps, which involve some $\bigo{10^4}$ terms and were done in Ref.~\cite{Abdallah:2021eii} using the tensor computer algebra suite \textsc{xAct}~\cite{xact,xpert}. Instead, we will show here the most relevant results;in addition, we will include a new detailed computation of the triangle diagram in Sec.~\ref{sec:triangle}.

Before doing so, it is important to mention that we have employed a modified Breitenlohner--Maison scheme for the parity-even sector~\cite{Jegerlehner:2000dz}: whenever we find a term containing an \emph{even} number of chiral matrices, we employ the usual relation $\{ \gamma_\mu, \gamma_* \} = 0$ for any $\mu$, assuming that the other properties of $\gamma_*$ are not jeopardized. The validity of this approach is justified by the compatibility between the Breitenlohner--Maison scheme and the quantum action principle~\cite{Breitenlohner:1977hr,Bonneau:1980ya,Rufa:1990hg}; indeed, spurious non-Lorentz-covariant terms that could arise if we had used the original Breitenlohner--Maison scheme also for the parity-even part can always be removed by appropriate finite counterterms. A clear understanding of this is provided by the corresponding cohomology theory, in particular in algebraic approaches~\cite{Bonora:1983ff,Piguet:1995er,Sibold:2000mp,Hollands:2007zg,Fredenhagen:2011mq,Frob:2018buw}; it is a classical result by Wigner~\cite{Wigner:1939cj} that the relevant cohomology class is empty for any finite-dimensional representation of the Lorentz group, and thus in particular for the representation in which the EM tensor transforms.

Beginning with the results at first order in $\kappa$, the regularized VEV for the EM tensor is traceless and divergenceless:
\begin{equation}
\label{eq:1_reg_trace_mod}
\mathcal{T}^\text{reg}_{(1)}(x) = 0 = \mathcal{D}^\nu_{\text{reg},\,(1)}(x) \eqend{.}
\end{equation}
On the other hand, the divergent term of the VEV reads
\begin{splitequation}
\label{eq:VEV1_div_mod}
\expect[div]{ T^{\mu\nu}(x) }_{(1)} &= - \frac{\diver}{120\cdot 16 \pi^2} \left( 6 \nabla^2 R^{\mu\nu} - 2 \nabla^\mu \nabla^\nu R - g^{\mu\nu} \nabla^2 R \right)_{(1)}(x) \eqend{,}
\end{splitequation}
where in our modified Minimal Subtraction ($\overline{\text{MS}}$) scheme we have defined\footnote{The quantity $\mu$ is the renormalization scale.}
\begin{equation}
\diver \coloneqq - \frac{2}{n-4} - \gamma - \log\left( \frac{\mu^2}{4\pi} \right) \eqend{.}
\end{equation}
Subtracting the divergent contributions from the regularized VEV, we can see that the renormalized VEV develops a nonvanishing trace but is still conserved:
\begin{equations}[eq:VEV1_tracediv_mod]
\mathcal{T}^\text{ren}_{(1)}(x) &= \frac{1}{60 (4\pi)^2} \left[ \nabla^2 R(x) \right]_{(1)} \eqend{,} \\
\mathcal{D}^\nu_{\text{ren},\, (1)}(x) &= 0 \eqend{.}
\end{equations}

At second order in $\kappa$, we separate the contributions according to the structures involved in the product of the expansions \eqref{eq:action_expansion} and \eqref{eq:emtensorexpansion}; as an example, we collect all the terms that come from a product of only two $\Psi$ in $\expect{T^{\mu\nu}}$ and denote it with a subindex $\Psi^2$:
\begin{equations}
\eta_{\mu\nu} \expect[reg]{ T^{\mu\nu}(x) }_{(2),\,\Psi^2} &= \frac{n-3}{2} h_{\alpha\beta}(x) \expect[reg]{ T^{\alpha\beta}(x) }_{(1)} \eqend{,} \label{eq:stress_trace_psi2} \\
\partial_\mu \expect[reg]{ T^{\mu\nu}(x) }_{(2),\Psi^2} &= \frac{1}{2} \partial_\mu \left[ h_{\alpha\beta}(x) \eta^{\mu\nu} \expect[reg]{ T^{\alpha\beta}(x) }_{(1)} - 3 h^{(\mu}{}_\alpha(x) \expect[reg]{ T^{\nu)\alpha}(x) }_{(1)} \right] \eqend{.} \label{eq:stress_div_psi2}
\end{equations}
The regularized contribution to the VEV of the EM operator arising from one $\Psi$ and one $J$ trivially vanishes, while the remaining contributions to the trace and the divergence,
originating from a $\Psi^3$ structure, read
\begin{equations}
\eta_{\mu\nu} \expect[reg]{ T^{\mu\nu}(x) }_{(2),\,\Psi^3} &= - \frac{n (n-1)}{8} h_{\rho\sigma}(x) \expect[reg]{ T^{\rho\sigma}(x) }_{(1)} \eqend{,} \label{eq:stress_trace_psi3} \\
\begin{split}
\partial_\mu \expect[reg]{ T^{\mu\nu}(x) }_{(2),\Psi^3} &= - \frac{1}{2} \partial_\rho h(x) \expect[reg]{ T^{\nu\rho}(x) }_{(1)} + \frac{3}{4} \partial^\rho \left[ h_{\rho\sigma}(x) \expect[reg]{ T^{\nu\sigma}(x) }_{(1)} \right] \\
&\quad- \frac{1}{4} \partial_\rho h_\sigma{}^\nu \expect[reg]{ T^{\rho\sigma}(x) }_{(1)} - \frac{1}{2} h_{\rho\sigma}(x) \partial^\nu \expect[reg]{ T^{\rho\sigma}(x) }_{(1)} \eqend{.} \label{eq:stress_div_psi3}
\end{split}
\end{equations}
Considering all these results together with the coefficients obtained in Eqs.~\eqref{eq:trace_definitions} and \eqref{eq:div_definitions}, we verify that the regularized VEV is traceless and divergenceless also at this order:
\begin{equations}[eq:tracediv_k2]
\begin{split}
\mathcal{T}^\text{reg}{}_{(2)}(x) &= \eta_{\mu\nu} \left[ \expect[reg]{ T^{\mu\nu}(x) }_{(2),\,\Psi^2} + \expect[reg]{ T^{\mu\nu}(x) }_{(2),\,\Psi J} + \expect[reg]{ T^{\mu\nu}(x) }_{(2),\,\Psi^3} \right] + h_{\mu\nu}(x) \expect[reg]{ T^{\mu\nu}(x) }_{(1)} \\
&= 0 \eqend{,} \raisetag{1.8em}
\end{split} \\
\begin{split}
\mathcal{D}^\nu_{\text{reg},\,(2)}(x) &= \partial_\mu \left[ \expect[reg]{ T^{\mu\nu}(x) }_{(2),\,\Psi^2} + \expect[reg]{ T^{\mu\nu}(x) }_{(2),\,\Psi J} + \expect[reg]{ T^{\mu\nu}(x) }_{(2),\,\Psi^3} \right] \\
&\qquad+ \frac{1}{2} \left[ 2 \partial_\rho h^\nu{}_\mu(x) - \partial^\nu h_{\rho\mu}(x) + \delta^\nu_\mu \partial_\rho h(x) \right] \expect[reg]{ T^{\mu\rho}(x) }_{(1)} = 0 \eqend{.}
\end{split}
\end{equations}

After some labour, we can extract the divergent contribution to the VEV of the EM tensor at second order; comparing with the expansions of geometrical quantities, we arrive to the conclusion that
\begin{splitequation}
\label{eq:VEV2_div}
\expect[div]{ T^{\mu\nu}(x) }_{(2)} &= \frac{\diver}{1440 (4\pi)^2} \Big[ 8 g^{\mu\nu} R_{\alpha\beta} R^{\alpha\beta} + 56 R^{\mu\alpha} R^\nu_\alpha + 20 R^{\mu\nu} R - 5 g^{\mu\nu} R^2 \\
&\hspace{6em}+ 7 g^{\mu\nu} R_{\alpha\beta\gamma\delta} R^{\alpha\beta\gamma\delta} - 88 R_{\alpha\beta} R^{\mu\alpha\nu\beta} - 28 R^{\mu\alpha\beta\gamma} R^\nu{}_{\alpha\beta\gamma} \\
&\hspace{6em}- 72 \nabla^2 R^{\mu\nu} + 12 g^{\mu\nu} \nabla^2 R + 24 \nabla^\mu \nabla^\nu R \Big]_{(2)} \eqend{,}
\end{splitequation}
which is conserved but not traceless. Subtracting it from the regularized value gives a divergenceless renormalized VEV for the EM tensor, whose trace corresponds to the following coefficients:
\begin{equation}
w_{W}^{(BM)} = \frac{18}{720 (4 \pi)^2} \eqend{,} \quad b_{W}^{(BM)} = - \frac{11}{720 (4 \pi)^2}\eqend{,} \quad c_{W}^{(BM)} = \frac{12}{720 (4 \pi)^2} \eqend{,} \quad f_{W}^{(BM)} = 0 \eqend{.}
\end{equation}
This is exactly half the result for the Dirac fermion, cf. Eq.~\eqref{eq:coefficients_Dirac}.

\subsection{Comment on the contribution of the triangle diagram}\label{sec:triangle}

Considering that in Refs.~\cite{Bonora:2014qla,Bonora:2017gzz} the nonvanishing parity-odd trace anomaly for a Weyl fermion is completely due to the triangle diagram, let us briefly highlight the main differences with our approach by performing a detailed computation. The relevant vertex can be read off from Eq.~\eqref{eq:actionexpansion_s1}; keeping in mind that the term proportional to $h^\mu{}_\mu$ in that expression cannot contribute to the odd sector of the trace anomaly, we have
\begin{equation}
\begin{tikzpicture}[baseline=(T.base),scale=1.1,every node/.style={transform shape}]
\begin{feynman}
  \vertex (ext) {};
  \vertex [right= of ext] (T);
  \vertex [above right= of T] (f1);
  \vertex [below right= of T] (f2);
  \vertex [below left=.5em of T] {\({\mu\nu}\)};
\diagram*[baseline=(T.base)]{
 (ext) -- [gluon] (T) -- [fermion, momentum'=\(p_1\)] (f2),
(f1) -- [fermion, momentum'=\(p_2\)] (T),
 };
 \end{feynman}
\end{tikzpicture}
= \frac{\mathi}{8}  \mathcal{P}_- \left[ (p_1+p_2)_\mu \gamma_\nu + (p_1+p_2)_\nu \gamma_\mu \right]  \mathcal{P}_+ \eqend{.}
\label{eq:feynman_rule}
\end{equation}

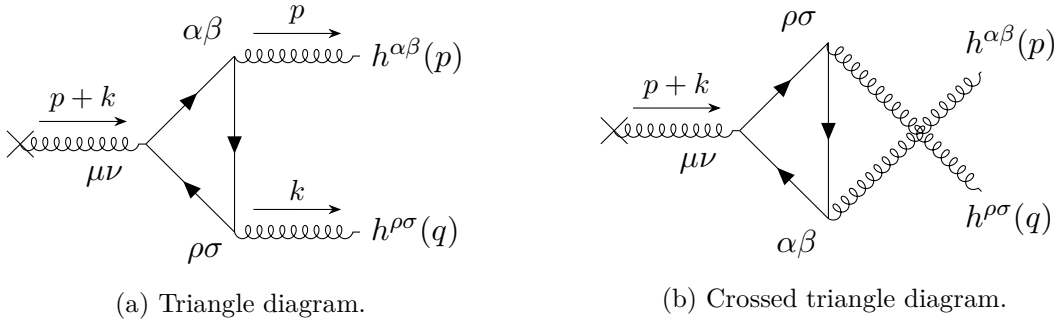
\begin{figure}
\centering

\begin{subfigure}[h!]{0.48\textwidth}
\centering
\begin{tikzpicture}[scale=1.1,every node/.style={transform shape}]
 \begin{feynman}
  \vertex (T) ;
  \vertex [right= of T] (f1);
  \vertex [above right= of f1] (f2);
  \vertex [below right= of f1] (f3);
  \vertex [right= of f2] (h1) {\(h^{\alpha\beta}(p)\)};
  \vertex [right= of f3] (h2) {\(h^{\rho\sigma}(q)\)};
 \vertex [below left=0.5em of f1] {\({\mu\nu}\)};
 \vertex [above left=0.0em of f2] {\({\alpha\beta}\)};
 \vertex [below left=0.0em of f3] {\({\rho\sigma}\)};
 
 \diagram*[]{
 (T) -- [gluon, momentum=\(p+k\), insertion={[size=5pt]0}] (f1) -- [fermion] (f2) -- [fermion] (f3) -- [fermion] (f1),
(f2) -- [gluon, momentum=\(p\)] (h1),
(f3) -- [gluon, momentum=\(k\)] (h2),
 };
 \end{feynman}
\end{tikzpicture}
\caption{Triangle diagram.}
\label{fig:triangle_one}
\end{subfigure}
\begin{subfigure}[h]{0.48\textwidth}
\centering
\begin{tikzpicture}[scale=1.1,every node/.style={transform shape}]
 \begin{feynman}
  \vertex (T) ;
  \vertex [right= of T] (f1);
  \vertex [above right= of f1] (f2);
  \vertex [below right= of f1] (f3);
  \vertex [right= of f2] (h1) {\(h^{\alpha\beta}(p)\)};
  \vertex [right= of f3] (h2) {\(h^{\rho\sigma}(q)\)};
 \vertex [below left=0.5em of f1] {\({\mu\nu}\)};
 \vertex [above left=0.0em of f2] {\({\rho\sigma}\)};
 \vertex [below left=0.0em of f3] {\({\alpha\beta}\)};
 
 \diagram*[]{
 (T) -- [gluon, momentum=\(p+k\),insertion={[size=5pt]0}] (f1) -- [fermion] (f2) -- [fermion] (f3) -- [fermion] (f1),
(f2) -- [gluon] (h2),
(f3) -- [gluon] (h1),
 };
 \end{feynman}
\end{tikzpicture}
\caption{Crossed triangle diagram.}
\label{fig:triangle_two}
\end{subfigure}
\caption{Triangle and crossed triangle diagrams built from the vertex defined in Eq.~\eqref{eq:feynman_rule}.
The conventions are the same as in Fig.~\ref{fig:generic_diagrams}.}
\label{fig:2}
\end{figure}

Using three of such vertices one can build a triangle diagram, which is depicted in Fig.~\ref{fig:triangle_one}; its contribution to the trace anomaly is given by
\begin{splitequation}
\eta^{\mu\nu} T_{\mu\nu\alpha\beta\rho\sigma} &= - \frac{\eta^{\mu\nu}}{256} \int \tr\bigg[ \frac{1}{\slashed{q}-\slashed{p}-\slashed{k}} \mathcal{P}_- (2q-2p-k)_{(\sigma} \gamma_{\rho)} \mathcal{P}_+ \\
&\hspace{4em}\times \frac{1}{\slashed{q}-\slashed{p}}  \mathcal{P}_- (2q-p)_{(\alpha} \gamma_{\beta)} \mathcal{P}_+ \frac{1}{\slashed{q}} \mathcal{P}_- (2q-p-k)_{(\mu} \gamma_{\nu)} \mathcal{P}_+ \bigg] \frac{\total^n q}{(2\pi)^n} \\
&= - \frac{\eta^{\mu\nu}}{256} \int \tr\Big[ \gamma_\delta \mathcal{P}_- \gamma_\rho \mathcal{P}_+ \gamma_\lambda \mathcal{P}_- \gamma_\beta \mathcal{P}_+ \gamma_\tau  \mathcal{P}_- \gamma_\nu \mathcal{P}_+ \Big] \frac{(q-p-k)^\delta}{(q-p-k)^2} \\
&\qquad\times (2q-2p-k)_\sigma \frac{(q-p)^\lambda}{(q-p)^2} (2q-p)_\alpha \frac{q^\tau}{q^2} (2q-p-k)_\mu \frac{\total^n q}{(2\pi)^n} \bigg\vert_{\text{sym}\{(\alpha\beta),(\rho\sigma)\}} \eqend{,}
\end{splitequation}
where the notation ``sym'' in the last line indicates that the expression needs to be symmetrized separately in the index pairs $\alpha\beta$ and $\rho\sigma$.

In the next step, we consider the trace of the gamma matrices included in this expression. Employing the commutation property of the projectors $\mathcal{P}_\pm$ with the gamma matrices~\eqref{eq:gamma_sandwich_projector}, the idempotency of the projectors and the cyclicity of the trace, all the gamma matrices in the trace become barred quantities, i.e., four-dimensional:
\begin{splitequation}
\label{eq:trace_gammas_triangle}
\tr\Big[ \gamma_\delta \mathcal{P}_- \gamma_\rho \mathcal{P}_+ \gamma_\lambda \mathcal{P}_- \gamma_\beta \mathcal{P}_+ \gamma_\tau \mathcal{P}_- \gamma_\nu \mathcal{P}_+ \Big] &= \tr\Big[ \mathcal{P}_+ \gamma_\delta \mathcal{P}_- \bar\gamma_\rho \bar\gamma_\lambda \bar\gamma_\beta \bar\gamma_\tau \bar\gamma_\nu \mathcal{P}_+ \Big] \\
&= \tr\Big[ \bar\gamma_\delta \bar\gamma_\rho \bar\gamma_\lambda \bar\gamma_\beta \bar\gamma_\tau \bar\gamma_\nu \mathcal{P}_+ \Big] \eqend{.}
\end{splitequation}
Following Ref.~\cite{Bonora:2017gzz}, we can recast $(2q-p-k)_\mu = q_\mu + (q-p-k)_\mu$; in doing so, we obtain two terms, one of which is symmetric under the exchange of the indices $\tau$ and $\mu$, and the remaining in $\mu$ and $\delta$. This enables us, upon commutation of $\gamma_\nu$ with $\mathcal{P}_+$ and using again the cyclicity of the trace, to simplify a pair of gammas in each contribution:
\begin{splitequation}
\label{eq:trace_triangle_one}
\eta^{\mu\nu} T_{\mu\nu\alpha\beta\rho\sigma} &= - \frac{1}{256} \int (2q-2p-k)_\sigma \frac{(q-p)^\lambda}{(q-p)^2} (2q-p)_\alpha \bigg\{ \tr\Big[ \bar\gamma_\delta \bar\gamma_\rho \bar\gamma_\lambda \bar\gamma_\beta \mathcal{P}_+ \Big] \frac{(q-p-k)^\delta}{(q-p-k)^2} \frac{\bar q^2}{q^2} \\
&\hspace{6em}+ \tr\Big[ \bar\gamma_\rho \bar\gamma_\lambda \bar\gamma_\beta \bar\gamma_\tau \mathcal{P}_- \Big] \frac{(\bar q-p-k)^{2}}{(q-p-k)^2} \frac{q^\tau}{q^2} \bigg\} \frac{\total^n q}{(2\pi)^n} \bigg\vert_{\text{sym}\{(\alpha\beta),(\rho\sigma)\}} \eqend{.}
\end{splitequation}

To further simplify the computation, let us focus on the parity-odd contributions, i.e., those containing a $\gamma_*$ in the trace; we will call $T'_{\alpha\beta\rho\sigma}(p,k)$ and $T''_{\alpha\beta\rho\sigma}(p,k)$ the results coming from the first and second terms in the right-hand side of Eq.~\eqref{eq:trace_triangle_one}. We are now going to show that they cancel with the parity-odd contributions of the crossed triangle diagram, i.e., the one in Fig.~\ref{fig:triangle_two}. To do so, notice that the crossed diagram can be obtained from the triangle diagram by swapping $(p,\alpha,\beta) \leftrightarrow (k,\rho,\sigma)$. Therefore, employing the explicit expression in Eq.~\eqref{eq:g5fourg_bm} for the trace of $\gamma_*$ with four other gamma matrices, one of the contributions of the crossed triangle diagram will be
\begin{splitequation}
&T^{'(c)}_{\alpha\beta\rho\sigma}(p,k) \Big\rvert_\text{odd} \coloneqq T^{'}_{\rho\sigma\alpha\beta}(k,p) \Big\rvert_\text{odd} \\
&\quad= \frac{\mathi}{512} \epsilon_{\delta\alpha\lambda\sigma} \tr \1 \int \frac{(2q-2k-p)_{\beta} (q-k)^{\lambda} (2q-k)_{\rho} (q-p-k)^{\delta} \bar{q}^2 }{(q-k)^2 (q-p-k)^2 q^2} \frac{\total^n q}{(2\pi)^n} \Bigg\rvert_{\text{sym}\{(\alpha\beta),(\rho\sigma)\}} \eqend{.}
\end{splitequation}
If we now perform the change of variables $q=-s+k+p$ and actively employ the symmetry in the pair of indices $\alpha\beta$ and $\rho\sigma$, we can conclude that 
\begin{splitequation}
&T^{'(c)}_{\alpha\beta\rho\sigma}(p,k) \Big\rvert_\text{odd} 
\\
&\quad= \frac{\mathi}{512} \epsilon_{\delta\beta\lambda\rho} \tr \1 \int \frac{(2s-p)_\alpha (s-p)^\lambda (2s-2p-k)_\sigma s^\delta (\bar{s}-p-k)^2}{(s-p)^2 (s-p-k)^2 s^2} \frac{\total^n s}{(2\pi)^n} \Bigg\rvert_{\text{sym}\{(\alpha\beta),(\rho\sigma)\}} \\
&\quad= - T''_{\alpha\beta\rho\sigma}(p,k) \Big\rvert_\text{odd} \eqend{.}
\end{splitequation}
An immediate consequence of this equality is that half of the contributions will cancel; for the remaining ones, it is straightforward to prove that
\begin{equation}
T^{''(c)}_{\alpha\beta\rho\sigma}(p,k) \Big\rvert_\text{odd} \coloneqq T^{''}_{\rho\sigma\alpha\beta}(k,p) \Big\rvert_\text{odd} = - T''_{\alpha\beta\rho\sigma}(p,k) \Big\rvert_\text{odd} \eqend{.}
\end{equation}
Thus, we have proved that the complete sum of the triangle and crossed triangle diagrams vanishes.

\section{Conclusions}\label{sec:conclusions}
We conclude this article with a few comments. First of all, it is not difficult to prove that our perturbative approach with Feynman diagrams, together with a dimensional regularization and the Breitenlohner--Maison prescription, is compatible with the well-established consistent chiral anomalies for Weyl fermions (see for example Ref.~\cite{Bastianelli:2022hmu}). This removes one of the more common criticisms against the methods that yield a vanishing parity-odd part of the trace anomaly.

Second, we have provided in Sec.~\ref{sec:triangle} an alternative proof of the fact that the parity-odd contribution of the triangle diagram vanishes. These computations allow us to perform a closer comparison with the results obtained in Refs.~\cite{Bonora:2014qla,Bonora:2017gzz}. From our discussion, it should be clear that the main differences are twofold: On the one hand, we do not use the four-dimensional commutator of $\gamma_*$ with other gamma matrices before regularizing the theory, so each computation contains several insertions of the projectors $\mathcal{P}_\pm$. On the other hand, in Ref.~\cite{Bonora:2017gzz} the trace of the EM tensor is taken in $n$ dimensions, see \cite[App. A, formula (195)]{Bonora:2017gzz}; instead, we have taken a four-dimensional trace. Even if we are forced to do so by Eq.~\eqref{eq:trace_gammas_triangle}, there is a stronger reason behind this choice: the trace anomaly is observed when taking the trace of the \emph{renormalized} EM tensor, which lives in a four-dimensional space. The renormalization process in a perturbative approach has been explained in detail, as a matter of completion, both for the parity-even and the parity-odd contribution in Ref.~\cite{Abdallah:2021eii}.

Third, in \cite{Duff:1993wm,Godazgar:2018boc} a definition of the trace anomaly is given for classically non-conformal theories,
\begin{equation}
\label{eq:trace_nicolai}
\mathcal{A} \coloneqq g^{\mu\nu} \expect{ T_{\mu\nu} } - \expect{ g^{\mu\nu} T_{\mu\nu} } \eqend{.}
\end{equation}
The importance of this equation is that the classical Weyl non-invariance is subtracted by the last term, opening the way for the study of quantum contributions in these cases. In our computations, the addition of the second term is superfluous, since we have explicitly proved the vanishing of the regularized trace. Instead, in Ref.~\cite{Bonora:2022izj} it is claimed that the nonvanishing result for the trace anomaly is exactly given by the last term in Eq.~\eqref{eq:trace_nicolai}. It is peculiar that a term which was introduced to cancel classical contributions in non-conformal theories could be responsible for generating the anomaly in a theory that can be regularized to satisfy Weyl invariance in arbitrary dimensions.

Fourth, in Ref.~\cite{Bonora:2022izj} a possible objection against a Feynman diagrammatic approach has been elaborated. The fact is that, in the framework of conformal field theory, it has been proven that the contribution to the three-point function $\expect{\Psi(x)\Psi(y)\Psi(z)}$ (involved in our triangle diagram) is constrained to be of even parity~\cite{Stanev:2012nq}. On first thought, one could argue that in our approach we are working in arbitrary dimensions, which might allow us to evade such a result; however, our definition of parity is intrinsically four-dimensional, rendering the issue rather delicate. Curiously, another algebraic obstruction was obtained in Ref.~\cite{Abdallah:2021eii}, where the odd part of the three-point function was shown to involve an antisymmetrization of five indices in four dimensions and thus trivially vanishes. In view of these results, the primary approach is to say that there is no need to renormalize the odd sector of the EM tensor at quadratic order in $\kappa$, and thus no parity-odd anomaly can arise. In Ref.~\cite{Bonora:2022izj}, a secondary approach has been put forward: there might be an ``anomaly in the parity-odd trace anomaly'', which prevents us from seeing the $\kappa^2$ contribution. While this seems somewhat far-fetched, it is a sort of chicken and egg dilemma, and to be fair we cannot exclude this possibility. Since, as far as we know, no such obstruction exists at higher orders, a laborious analysis at order $\bigo{\kappa^3}$ might serve to elucidate this point. However, no such objection stands against other, non-perturbative methods to compute the anomaly, namely the Pauli--Villars regularization~\cite{Bastianelli:2016nuf, Bastianelli:2019zrq} and the Hadamard approach~\cite{Frob:2019dgf}.

Fifth, a case has been made for the existence of a parity-odd part in the trace anomaly by making reference to the index theorem~\cite{Bonora:2022izj}. The widely known index theorems relate the zero modes of elliptic differential operators defined on generic Riemannian manifolds, e.g. the Laplacian operator, to topological invariants thereof~\cite{Atiyah:1963zz}. In our scenario, the special role of the zero modes can be seen as follows: The one-loop effective action is given by the functional determinant of the equation-of-motion operator for the quantum fluctuations that have been integrated out, which is a differential operator. One needs to consider the zero modes separately when computing the determinant, which can then lead to anomalies. Christensen and Duff~\cite{Christensen:1978md} have analyzed the small proper time expansion of heat kernels for second-order elliptic differential operators for (properly gauge-fixed) fields of arbitrary spin. In particular, they have computed the $b_{n/2}$ coefficient, which is related to the zero modes of the operators and Weyl anomalies: for a Weyl fermion they do indeed obtain a non-vanishing contribution involving the parity-odd Pontryagin density.

However, the caveats are twofold: first, the results of Christensen and Duff~\cite{Christensen:1978md} were derived in the Euclidean case, that is for an elliptic operator defined on a Riemannian manifold, not for a hyperbolic differential operator on a pseudo-Riemannian one. Only very recently have index theorems been derived in the Lorentzian case~\cite{baer_strohmaier_2015,baer_strohmaier_anomaly,baer_strohmaier_2022}; while they share many similarities with their Riemannian counterparts, there are also subtle differences. 
Second, the Wick rotation for fermions, and in particular for chiral Weyl fermions, is a complicated subject and, to our knowledge, there are only two consistent possibilities: on the one hand, a continuous Wick rotation~\cite{Mehta:1986mi,vanNieuwenhuizen:1996tv}, which has to be performed separately for spinors and cospinors, such that they end being completely independent in Euclidean space and the chirality of the cospinors is changed. Thus, with this prescription, the Wick-rotated EM tensor of chiral fermions couples right- and left-handed Weyl fermions equally, and consequently the parity-odd terms in the trace anomaly cancel since they have opposite sign for left- and right-handed Weyl fermions; this is of course consistent with our results. On the other hand, one can perform an analytic continuation of the vielbein and keep the $\gamma$ matrices unchanged~\cite{Wetterich:2010ni}, such that also in Euclidean space left-handed spinors are coupled to left-handed cospinors, and a chiral EM tensor is Wick-rotated into a chiral EM tensor in Euclidean space. However, the resulting Euclidean spinors do not transform properly under Euclidean rotations; one can rectify this issue by considering a different complex structure for spinors~\cite{Kupsch:1988sb,Wetterich:2010ni}, which then, however, breaks chiral invariance even classically. Therefore, one needs to be even more careful when computing chiral anomalies with this second prescription.

Let us mention in conclusion that this discussion is probably not settled; we encourage other groups to devote some time and their knowledge to tackle this problem, which may have several relevant consequences in the understanding of our universe, as discussed in the introduction. Just be careful not to end up trapped in the spiderweb of the anomaly.

\ack
The authors are grateful to F.~Bastianelli, L.~Bonora, C.~Corian{\`o}, M.~Chernodub, I. Moss and Y.~Nakayama for helpful discussions. SAF acknowledges the support from Helmholtz-Zentrum Dresden-Rossendorf (HZDR), PIP 11220200101426CO Consejo Nacional de Investigaciones Cient\'ificas y T\'ecnicas (CO\-NI\-CET) and Project 11/X748, UNLP. MBF acknowledges the support by the Deutsche Forschungsgemeinschaft (DFG, German Research Foundation), project No.~396692871 within the Emmy Noether grant~CA1850/1-1 and project No.~406116891 within the Research Training Group RTG~2522/1.

\bibliography{traceanom}

\end{document}